\documentclass{jfm}

\usepackage{graphicx}
\usepackage{natbib}

\ifCUPmtlplainloaded \else
  \checkfont{eurm10}
  \iffontfound
    \IfFileExists{upmath.sty}
      {\typeout{^^JFound AMS Euler Roman fonts on the system,
                   using the 'upmath' package.^^J}%
       \usepackage{upmath}}
      {\typeout{^^JFound AMS Euler Roman fonts on the system, but you
                   dont seem to have the}%
       \typeout{'upmath' package installed. JFM.cls can take advantage
                 of these fonts,^^Jif you use 'upmath' package.^^J}%
      }
  \else
  \fi
\fi

\ifCUPmtlplainloaded \else
  \checkfont{msam10}
  \iffontfound
    \IfFileExists{amssymb.sty}
      {\typeout{^^JFound AMS Symbol fonts on the system, using the
                'amssymb' package.^^J}%
       \usepackage{amssymb}%
       \let\le=\leqslant  
         
      }{}
  \fi
\fi

\ifCUPmtlplainloaded \else
  \IfFileExists{amsbsy.sty}
    {\typeout{^^JFound the 'amsbsy' package on the system, using it.^^J}%
     \usepackage{amsbsy}}
    {}
\fi

\newcommand\Real{\mbox{Re}} 
\newcommand\Rey{\mbox{\textit{Re}}}  
\newcommand\Pen{\mbox{\textit{Pe}}}  
\newsavebox{\astrutbox}
\sbox{\astrutbox}{\rule[-5pt]{0pt}{20pt}}

\newcommand{\evec}{\textit{\textbf{e}} }
\newcommand{\vvec}{\textit{\textbf{v}} }

\newcommand{\Jvec}{\textit{\textbf{J}} }
\newcommand{\zvec}{\textit{\textbf{z}} }

\newcommand{\Amat}{\textsf{\textbf{A}} }
\newcommand{\Bmat}{\textsf{\textbf{B}} }
\newcommand{\Tmat}{\textsf{\textbf{T}} }

\newcommand\eg{e.g.}

\title[Instabilities of Brownian suspensions]{Long-lived and unstable modes of Brownian suspensions in microchannels}

\author[Atefeh Khoshnood and Mir Abbas Jalali]%
{A\ls T\ls E\ls F\ls E\ls H \ns K\ls H\ls O\ls S\ls H\ls N\ls O\ls O\ls D$^{1,2}$%
\and
M\ls I\ls R\ns A\ls B\ls B\ls A\ls S\ns  J\ls A\ls L\ls A\ls L\ls I$^1$
 \thanks{Email address for correspondence: mjalali@sharif.edu}}

\affiliation{$^1$Computational Mechanics Laboratory, Department of Mechanical 
Engineering, \\ Sharif University of Technology, Azadi Avenue, P.O. Box: 11155-9567, Tehran, Iran\\[\affilskip]
$^2$Department of Mechanical and Aerospace Engineering, Princeton University, \\
Princeton, NJ 08544-5263, USA}

\pubyear{2012}
\begin{document}

\maketitle

\begin{abstract}
We investigate the stability of the pressure-driven, low-Reynolds flow of Brownian suspensions 
with spherical particles in microchannels. We find two general families of stable/unstable 
modes: (i) degenerate modes with symmetric and anti-symmetric patterns; 
(ii) single modes that are either symmetric or anti-symmetric. 
The concentration profiles of degenerate modes have strong peaks near the channel 
walls, while single modes diminish there. Once excited, both families would be detectable 
through high-speed imaging. We find that unstable modes occur in concentrated suspensions 
whose velocity profiles are sufficiently flattened near the channel centreline. The patterns 
of growing unstable modes suggest that they are triggered due to Brownian migration of 
particles between the central bulk that moves with an almost constant velocity, and 
highly-sheared low-velocity region near the wall. Modes are amplified because 
shear-induced diffusion cannot efficiently disperse particles from the cavities of the 
perturbed velocity field. 
\end{abstract}


\section{Introduction}

Microfluidic devices operate in overwhelming laminar conditions where particles migrate across 
streamlines either through Brownian motion or shear-induced diffusion (SID). In microfiltration, 
a process to remove unwanted particles from fluid using a membrane, SID by very strong crossflow 
is essential to lessen the growth of particle layer cake over the membrane \citep{vol10}. 
Particles with different sizes are segregated in particle fractionation devices 
due to SID \citep{Kromkamp06}. Ceramic or metallic particle injection moulding is another subject 
for SID to play a role in \citep{kau11}. While Brownian random walk due to thermal fluctuations 
operates in all times, the efficiency of shear-induced migration depends on particle-particle 
and particle-fluid interactions. For instance, spherical particles move to regions with lower shear 
rates when the P\'eclet number is sufficiently large \citep{Sem07,Sem08}, but platelike particles 
do not sense the details of velocity profile and respond only to average shear rate \citep{RS08}. 
The lowest reported volume fraction that supports SID is $\lesssim 0.04$ \citep{brown09}. 
Well below this limit, SID is turned off as the experiments of \citet{RS08} show no transfer of spherical 
particles from the sample to buffer stream of a T-sensor. Interesting and largely unexplored dynamics 
emerges when particle migration is governed not by a single mechanism, but through the interplay 
between Brownian motion and SID. 

Despite the apparent stability of low-Reynolds flows, some transient substructures like ripples 
and sharp near-wall features of concentration profiles \citep[see][]{Fra03,Sem07} are observed in 
experimental data. Substructures can be long-lived or unstable `modes' excited by anomalous 
particle migrations, wall roughness, and particle-wall interactions. Instabilities in microchannels, 
however, are very hard to detect experimentally, and their theoretical prediction is a challenging 
problem because particle migration is generally a slow process compared to the time scale of 
velocity fluctuations. Moreover, none of the three known instabilities induced by 
interfaces \citep{Helton07}, massive sedimentation \citep{Yia06,Rao07}, 
and gravity \citep{GNR01,Carpen02} seem to occur in microchannel flows whose streaming 
velocity profiles are symmetric with respect to the channel centreline. Important questions regarding 
the flow of suspensions in microchannels thus include: (i) if excited, how long can stable modes 
survive, and in what flow conditions are they detectable? (ii) what are the shapes of long-lived 
substructures and how do they depend on dimensionless Reynolds and P\'eclet numbers? 
(iii) How do Brownian diffusion and SID compete in the bulk and near the walls, and when do 
they collaborate to destabilize suspension flows? In this paper we attempt to answer these 
questions, which have remarkable implications for the design and manufacturing of 
microfluidic devices.

Dynamics of suspension flows is described by  different models. The first model introduced by 
\citet{Leighton87} and \citet{Phil92} is phenomenological and involves the diffusion fluxes of 
particles due to particle collisions and the spatial variation in the viscosity. The second model 
roots from the conservation equations of mass, momentum and energy for the fluid and particle
phases \citep{not94}. \citet{MB99} take into account normal stress differences to handle curvilinear
flows. In this paper we adopt the constitutive model of \citet{Phil92} for two reasons: (i) combining 
the effects of Brownian and shear-induced diffusions is a straightforward superposition (ii) The 
free diffusion flux parameters allow for an exploration of different flow regimes. 
The literature also includes models where particle and fluid phases interact only through 
Stokes drag \citep[e.g.,][and references therein]{R97,KLB11}. 
However, our study is different in nature: the Reynolds number is smaller than these studies
by several orders of magnitude, and despite a strong coupling between the particle and fluid 
motions, particles undergo direct two-body collisions governed by shear and 
viscosity gradients. These collisions lead to particle phase pressure and shear overlooked 
in the models of \citet{R97} and \citet{KLB11}.

We assume that the mean streaming velocities of particles and the solvent are identical, i.e., 
the slip velocity is zero because the drag force is high. We include the Brownian diffusion in the flux vector. 
This leads to new steady-state concentrations that do not saturate at the centre of the channel. 
We briefly review the governing equations of suspension flows in \S\ref{sec:governing-equations}
and find their steady-state solutions in the presence of Brownian diffusion. We linearly perturb 
the diffusion and momentum equations in the vicinity of steady state solutions and utilise the 
Chebyshev tau method to determine the eigenmodes of perturbed equations. We present 
the results of our modal analysis in \S\ref{sec:long-lived-and-unstable-modes} and explain 
the physical mechanism of a new instability that emerges in concentrated suspension 
flows.

\section{Governing diffusion and momentum equations}
\label{sec:governing-equations}

We aim at modeling the diffusion inside a microchannel of the width $2W$ as shown in figure \ref{fig1}({\em a}). 
We confine our study to regions far from the edges where the flow has a two-dimensional nature in the 
$(x,z)$ plane, which is spanned by the unit vectors $\evec_x$ and $\evec_z$. The $x$ and $z$ axes are 
along the channel width and flow direction, respectively. We define the mean streaming velocity as 
$\vvec=v_x \evec_x+ v_z \evec_z$ and assume that streaming field remains invariant by changing the 
$y$-coordinate. This is a legitimate assumption because SID is controlled by shear gradient in the 
shortest direction. We define $\phi$ as the actual concentration of particles, and set its maximum 
achievable value to $\phi_m=0.68$. We scale all lengths by $W$ and all velocities by the maximum 
velocity $V_{p}$ of the associated Poiseuille flow when $\phi = 0$. Physical quantities are therefore 
normalized as $\bar \phi=\phi/\phi_m$, $(\bar x,\bar z)=(x/W,z/W)$, $(\bar v_x,\bar v_z)=(v_x/V_p,v_z/V_p)$ 
and $\bar t=V_p t/W$, where $t$ is the actual time and $-1 \le \bar x \le +1$. From here on, we will 
drop the bar sign for brevity and will explicitly mention if we use actual values.

For a flow with the mean streaming velocity $\vvec$, the volume fraction $\phi$ evolves 
according to the following nonlinear partial differential equation (PDE) \citep{Phil92}
\begin{eqnarray}
&{}& \!\!\!  \phi_{,t} + (\vvec \cdot {\bf \nabla})\phi = - \epsilon \nabla \cdot \Jvec,  ~~
\epsilon= \phi_m K_c ( a / W )^2, \label{eq:nonlinear-diffusion-1} \\
&{}& \!\!\! \Jvec = -   \phi {\bf \nabla} \left ( \Gamma \phi \right )  - 
\beta  \phi^2 \Gamma \eta^{-1} \eta_{,\phi} {\bf \nabla} \phi  - D \nabla \phi,~~ 
D = D_0 W/(\phi_{m} K_c a^2 V_{p}),
\nonumber
\end{eqnarray}
where $\Jvec$ is the flux of particles, $\eta(\phi)= (1-\phi)^{-\alpha}$ (with $\alpha=1.82$) is the relative 
viscosity of the suspension, and 
\begin{eqnarray}
\Gamma=\left [ \vert 4v_{x,x} v_{z,z}-v_{x,z}^2-v_{z,x}^2-2v_{x,z}v_{z,x}  \vert \right ]^{1/2},
\end{eqnarray}
is the magnitude of the local shear rate. $\Gamma$ is the second invariant of the rate of 
strain tensor. Throughout this paper, $(.)_{,s}$ denotes the partial differentiation operator 
$\partial(.)/\partial s$. $a$ is the typical radius of spherical particles in the suspension, and 
$D_0=k_B T/(6\pi \eta_s a)$ is the coefficient of Brownian diffusion where $\eta_s$ and $T$ 
are, respectively, the solvent viscosity and temperature. $k_B$ is Boltzmann's constant.
The coefficients of diffusion fluxes $K_c$ and $K_{\eta}=\beta K_c$ are `phenomenological' 
constants. $K_c$ and $K_\eta$ indicate the strength of two-body interactions due to the 
spatial variations of collision frequency and suspension viscosity, respectively. In the absence 
of Brownian diffusion, \citet{mer05} experimented two flow geometries, parallel-plate and 
Couette flows, and concluded that $\beta \gtrsim 1$ is independent of flow geometry. 
Although they use $\beta=2.1$ to fit the steady concentration profiles in both geometries, 
their numerical results are satisfactory only for the Couette flow. The model of \citet{Phil92} 
performs well for Poiseuille flow studied here \citep[][section 5]{SP05}. 

One can define the intrinsic time scale $t_B$ of the suspension flow based on the Brownian 
motion of particles across the channel width described by $\langle x^2 \rangle=4 D t$. 
Since $-1 \le  x \le +1$, we set the mean square displacement $\langle x^2 \rangle$ to $2^2$ 
and define $t_B=\langle x^2 \rangle/(4D)=1/D$. In \S\ref{sec:perturbed-equations}, $t_B$ will 
be used to quantify the period of transient oscillations.  
For suspensions, the elements of the stress tensor $\Tmat=[T_{ij}]$ are given as
\citep[][equation 2.4{\em b}]{Carpen02}
\begin{eqnarray}
T_{ij}=-\delta_{ij} p+\eta(\phi) \left [ {\bf \nabla}\vvec + \left ( {\bf \nabla}\vvec \right )^{\rm T} \right ],
\label{eq:define-stress}
\end{eqnarray}  
where the superscript T denotes transpose. The suspension pressure $p$ is a superposition of 
the solvent and particle-phase pressures. Equation (\ref{eq:define-stress}) is obtained from the 
stress tensor of rheological models \citep[e.g.,][]{YM08} by neglecting differences between normal 
stresses. This assumption is valid when the migration of particles occurs in the shear plane, as
in the straight channels studied here. 

Defining $\Rey=\rho W V_{p}/\eta_s$ as the channel 
Reynolds number in the limit of $\phi \to 0$, the normalized continuity and momentum equations 
read 
\begin{eqnarray}
{\bf \nabla}\cdot \vvec  =  0, ~~
{\bf \nabla}\cdot \Tmat  =   \Rey \left [ \vvec_{,t} + 
\left ( \vvec \cdot {\bf \nabla} \right ) \vvec \right ].
\label{eq:normal-momentum} 
\end{eqnarray}
Since $\Rey$ is very small ($10^{-2}\lesssim \Rey \lesssim 1$) in most microchannel experiments \citep{Sem07,RS08}, 
it is reasonable to work with a constant suspension density $\rho$ and drop terms like $\phi \, \Delta \rho/\rho$ 
where $\Delta \rho$ is the difference between the particle and fluid phase densities. After solving 
(\ref{eq:normal-momentum}) for the velocity field, the particle P\'eclet number is determined as 
$\Pen=a^2 V_p v_{z,{\rm max}}/(D_0 W)$.

\subsection{Steady-state solutions}
\label{sec:steady-state}

In a steady-state fully developed flow, the gradients of physical quantities are nonzero only 
in the $x$ direction, and the maximum velocity occurs at $x=0$ (channel centreline). 
The steady solutions $\phi=\phi_0(x)$, $v_{x}=0$ and $v_z=v_{z0}(x)$ 
of equations (\ref{eq:nonlinear-diffusion-1}) and (\ref{eq:normal-momentum}) are obtained by 
setting ${\bf \nabla}\cdot \Tmat=0$ and $\Jvec=0$. The first relation gives $\Gamma_0=2|x|/\eta_0(x)$, 
and the latter yields the first-order ordinary differential equation 
\begin{eqnarray}
\frac{d\phi}{dx} \! = \! -2\, {\rm sign}(x) \phi^2(1-\phi)^\alpha \left [D \!+\! 2|x| \phi(1-\phi)^{\alpha} \!+\! 
2 \alpha(\beta-1) |x| \phi^2(1-\phi)^{\alpha -1} \right ]^{-1},
\label{eq:steady-state-implicit}
\end{eqnarray}
where $\eta_0(x)=\eta(\phi_0(x))$. We numerically integrate equation (\ref{eq:steady-state-implicit}) and 
obtain $\phi=\phi_0(x)$. The corresponding profile of $v_{z0}(x)$ is then determined from
$v_{z0}(x)=\int_{-1}^{x} 2 \xi [\eta_0(\xi)]^{-1} d\xi$ with the boundary condition $v_{z0}(\pm 1)=0$.
Our numerical experiments show that for large values of $\beta$ the concentration $\phi_0(x)$ develops
a cusp as $x\to 0^{\pm}$ and its tails become flat for $|x|\to 1$. The velocity profile is flattened near the 
channel centreline by decreasing $\beta$. The profile of $v_{z0}(x)$ approaches to $1-x^2$ ($v_{z,{\rm max}} \to 1$) 
as $\phi \to 0$. The physical values of $K_c$ and $\beta$ are determined through fitting the computed profiles 
of $\phi_0(x)$ and $v_{z0}(x)$ to experimental data. In this paper we explore the perturbations of 
models with $1 \lesssim \beta \lesssim 5$. In the limit $D=0$ the steady solutions are obtained from 
$x \phi_0(x) \eta^{\beta-1}=$ constant \citep{Phil92} while $\phi_0(0)$ is always saturated to 1. 
We are not interested in this extreme case.

\begin{figure}
\centerline{\hbox{\includegraphics[width=0.4\textwidth]{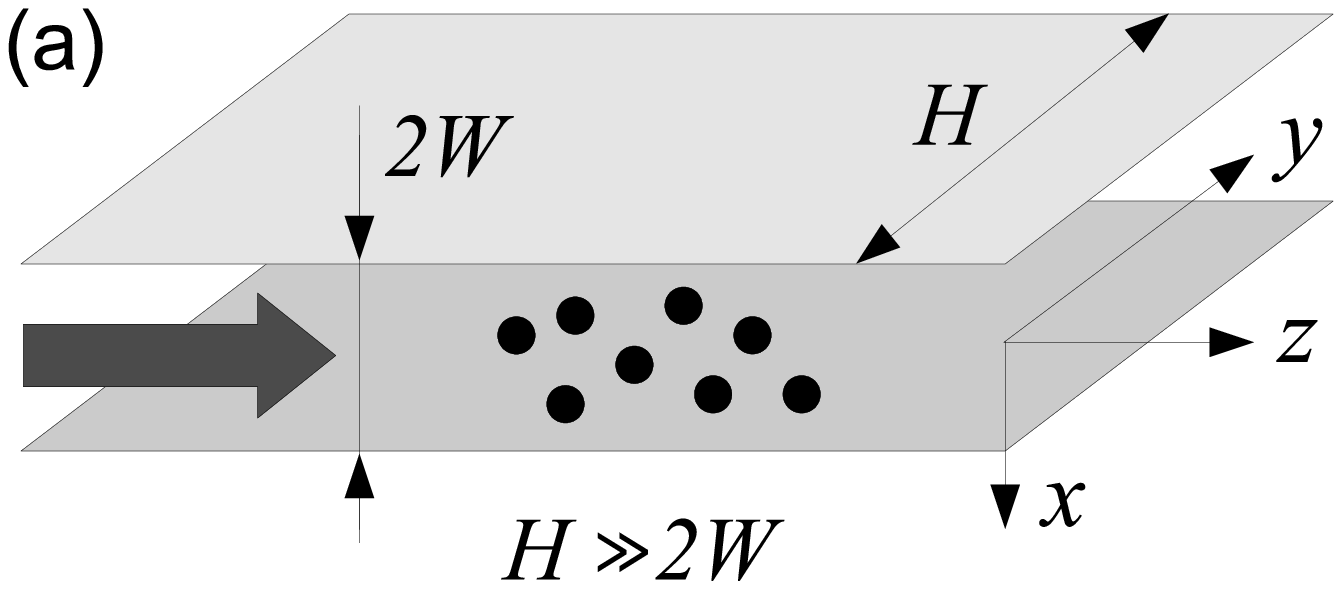}} 
              \hspace{6 mm} \hbox{\includegraphics[width=0.45\textwidth]{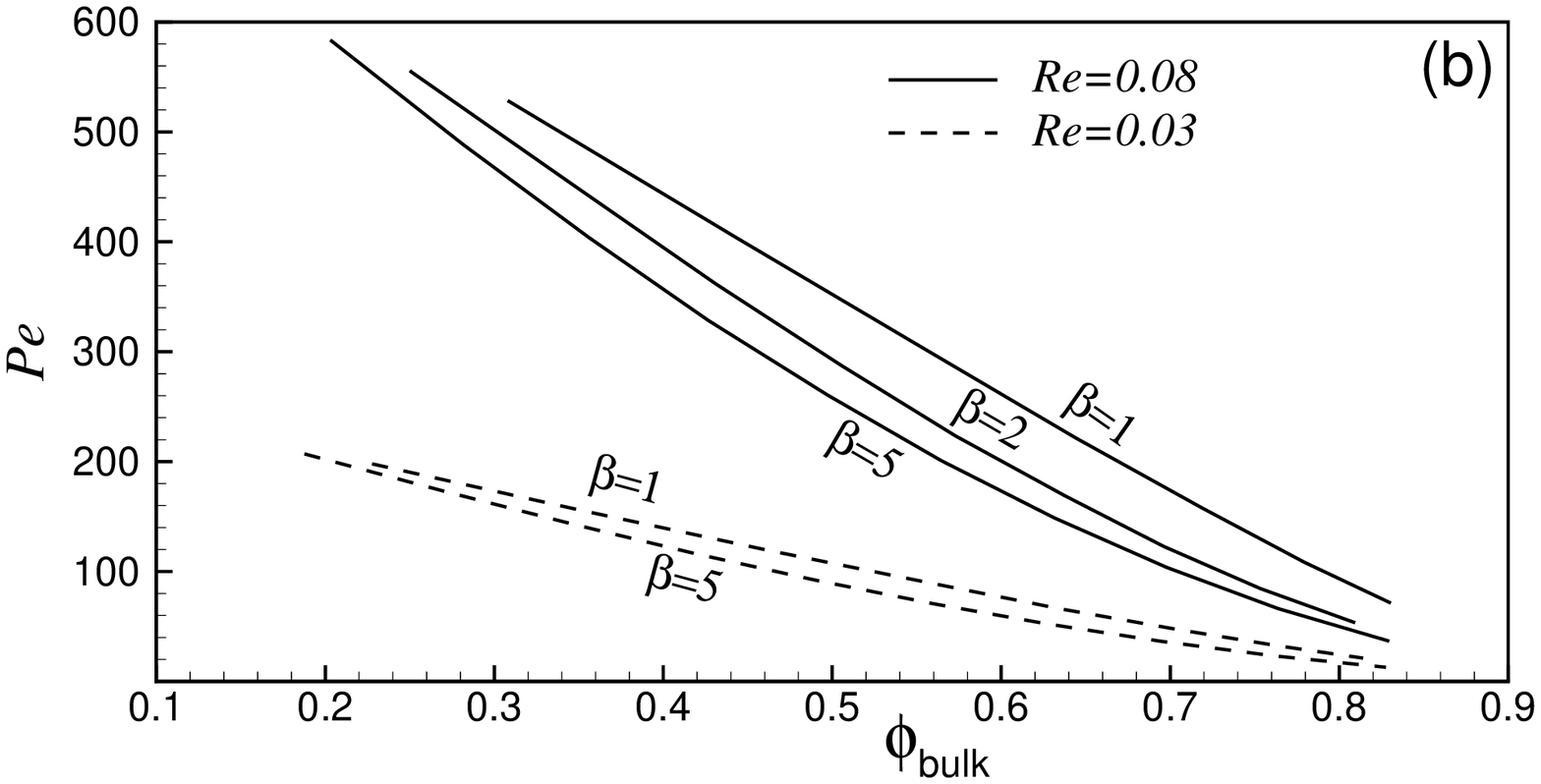}}  }
              \vspace{1 mm}
\centerline{\hbox{\includegraphics[width=0.45\textwidth]{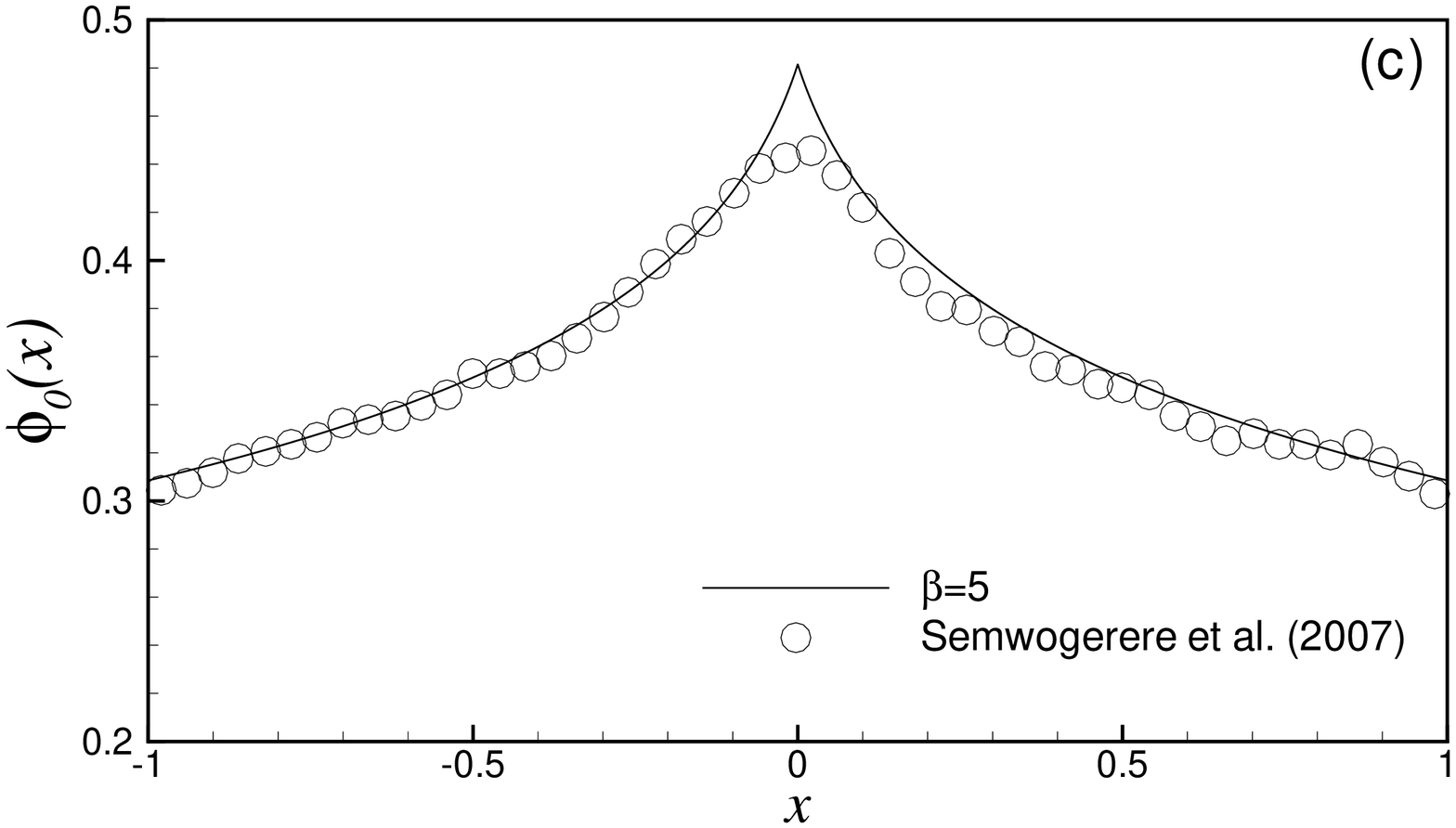}}
             \hbox{\includegraphics[width=0.45\textwidth]{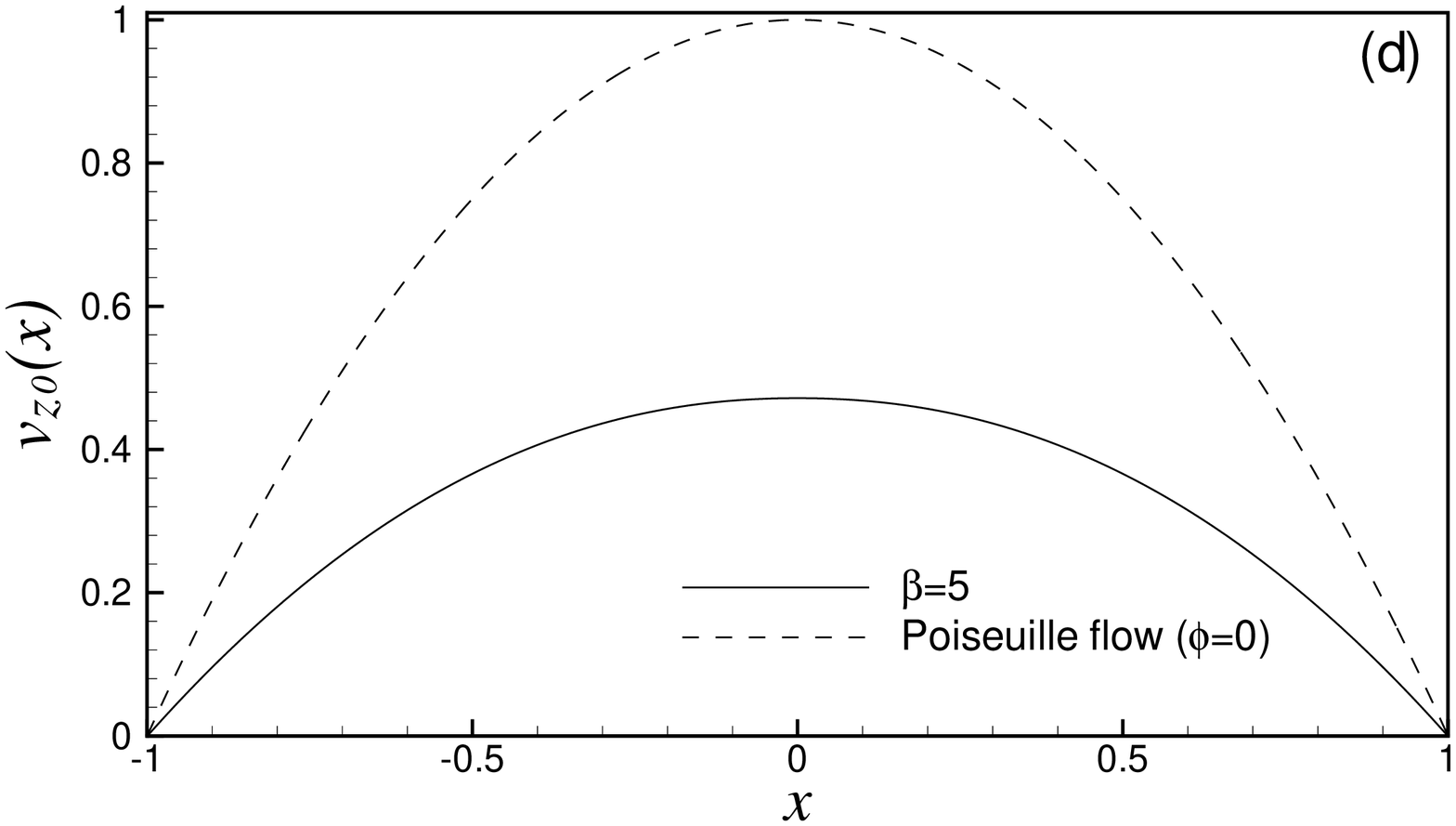} }	}
\caption{(a) Channel geometry; big arrow indicates flow direction. (b) Variation of $\Pen$ versus $\phi_{\rm bulk}$ 
for several choices of $\Rey$ and $\beta$. We have set $K_c=0.03$. (c) Steady-state solution of the normalized particle 
concentration $\phi$ for $\beta=5$. Circles show the experimental results of \citet[][figures 3]{Sem07}. 
(d) Solid line: the velocity profile $v_{z0}(x)$ corresponding to $\phi_0(x)$; dashed line: the velocity profile of 
Poiseuille flow in the limit of $\phi=0$. Model parameters are $K_c=0.03$, $\phi_0(\pm 1)=0.21/\phi_m$, 
$W=25\mu{\rm m}$, $a=0.7\mu{\rm m}$, $V_{p}=2.1\times 10^{-3}$m/s, $T=22^{\circ}$, 
$\eta_s=2.18\times 10^{-3}$Pa.s, $\rho=1232.5$kg/m$^3$ and $D=8.4\times 10^{-2}$. 
The actual velocity is computed as $V_p v_{z0}(x)$. Using these parameters, we obtain $Q=16.92$ nl/s, 
$\Pen=137$ and $\phi_{\rm bulk}=0.25/\phi_m$.}
\label{fig1}
\end{figure}

We have used the solution of (\ref{eq:steady-state-implicit}) and plotted $\Pen$ versus $\phi_{\rm bulk}$ in 
figure \ref{fig1}({\em b}) for several choices of $\Rey$ and $\beta$. 
Figures \ref{fig1}({\em c}) and \ref{fig1}({\em d}) show the steady-state solutions of 
a suspension flow with $\beta = 5$. The material and geometrical parameters---given in the figure caption---of this 
example come from \citet{Sem07} for the solvent cyclohexylbromide/decalin mixture. Using an evolution parameter, 
they have confirmed the steady state condition of the flow. We find $\phi_{\rm bulk} = 0.25/\phi_m$ and $\Pen = 137$, 
which agree with the experimental data within 5\% \citep[see][figures 3 and 10{\em c}]{Sem07}. The solid line in 
figure \ref{fig1}({\em c}) accurately reproduces the concentration profile in figure 10({\em c}) of \citet{Sem07} 
whose own analytical predictions \citep[see also][]{MB99,MM06} show remarkable deviations from the 
measurements at the fully developed stage. This can be due either to electrical stresses, or particle 
random walks. The impressive match between the results of the steady-state model (\ref{eq:steady-state-implicit}) 
and experimental data is mainly due to $D\not =0$ and supports the latter possibility. 
The significance of having a non-zero $D$ had not already been discussed/explored in the 
original work of \citet{Phil92} because they worked with very large P\'eclet numbers of ${\cal O}(10^5)$.

\subsection{Perturbed equations and eigenvalue problem}
\label{sec:perturbed-equations}

The steady concentration of particles in microchannels can easily be disturbed. For instance, unavoidable 
surface roughness, sedimentation, particle-wall and particle-particle interactions near a wall can induce 
small amplitude fluctuations on the boundary values of $\phi$ and generate global modes. To understand 
the transient response of suspensions, we carry out a linear stability analysis by perturbing the concentration 
and velocity fields as $\phi = \phi_0(x) + \tilde \phi(x,z,t)$ and $\vvec = v_{z0}(x)\evec_z+\tilde \vvec(x,z,t)$.
The magnitude of the shear rate and the normalized viscosity then become 
$\Gamma = \Gamma_0(x)+\tilde \Gamma(x,z,t)$ and $\eta = \eta_0(x)+\eta_{,\phi}(\phi_0(x))\, \tilde \phi$,
where $\tilde \Gamma = {\rm sign}\left ( \partial v_{z0}/\partial x \right )\left ( \tilde v_{x,z}+\tilde v_{z,x} \right )$.
It is remarked that the transient response can develop structures in the $y$-direction as well. Since the steady-state 
quantities do not depend on $y$, the linear response in that direction will include a simple harmonic waveform 
with a wavelength $\ell_y$. The magnitudes of all gradients in the $y$-direction are determined by $1/\ell_y$. 
This study is conducted in the long wavelength limit $\ell_y \to \infty$.

The perturbed equations are simplified by assuming the stream function $\psi=\psi_0(x)+\tilde \psi(x,z,t)$ 
to express the velocity field: $\vvec=(\psi_{,z},-\psi_{,x})$. This implies $\vvec_0(x)=(0,-\psi_{0,x})$ and  
$\tilde \vvec=(\tilde \psi_{,z},-\tilde \psi_{,x})$, and the continuity equation is satisfied automatically 
both in the steady and perturbed states. We now take the curl of (\ref{eq:normal-momentum})
and remove the pressure $p$ from computations. The resulting equation together with (\ref{eq:nonlinear-diffusion-1}) 
are linearized to obtain:
\begin{eqnarray}
{\cal L}_{11} \tilde \psi + {\cal L}_{12} \tilde \phi &=& \Rey \left (\nabla^2 \tilde \psi_{,t} + \psi_{0,xxx} \tilde \psi_{,z} -
\nabla^2 \tilde \psi_{,z} \right ), \label{eq:perturbed-psi-1} \\
-\epsilon \nabla \cdot (\tilde J_x \evec_x+\tilde J_z \evec _z) &=&
\tilde \phi_{,t} + \phi_{0,x} \tilde \psi_{,z}-\psi_{0,x} \tilde \phi_{,z}, \label{eq:perturbed-psi-2}
\end{eqnarray}
where the perturbed components of the flux vector defined as $\tilde J_x = {\cal L}_{21} \tilde \psi + {\cal L}_{22} \tilde \phi$ 
and $\tilde J_z = {\cal L}_{31} \tilde \psi + {\cal L}_{32} \tilde \phi$. The linear operators ${\cal L}_{ij}$ are functions 
of $\psi_0(x)$, $\phi_0(x)$ and their $x$-derivatives. They are obtained by evaluating
\begin{eqnarray}
\begin{array}{lll}
{\cal L}_{11} = \left ( \nabla \times \nabla \cdot \Tmat  \right )_{,\tilde \psi},  &    
{\cal L}_{12} = \left ( \nabla \times \nabla \cdot \Tmat  \right )_{,\tilde \phi},  &
{\cal L}_{21} = \left (  \Jvec \cdot \evec_{x} \right )_{,\tilde \psi}, \\ 
{\cal L}_{22} = \left (  \Jvec \cdot \evec_{x} \right )_{,\tilde \phi},  &
{\cal L}_{31} = \left (  \Jvec \cdot \evec_{z} \right )_{,\tilde \psi},  &  
{\cal L}_{32} = \left (  \Jvec \cdot \evec_{z} \right )_{,\tilde \phi},
\end{array}
\label{eq:L-operators}
\end{eqnarray}
at $(\tilde \psi,\tilde \phi)=0$. The partial derivatives $\partial/\partial \nu$ and $\partial/\partial g$ 
are noncommutative over the extended space $(\nu,g)$ when $\nu\equiv (x,z,t)$ and $g\equiv (\tilde \psi,\tilde \phi)$ 
are independent and dependent variables, respectively. We have applied the rule 
$[f(\nu)g(\nu)]_{,\nu g}=f\partial/\partial \nu+\partial f/\partial \nu$ 
to calculate the partial derivatives in (\ref{eq:L-operators}); i.e., we first differentiate with respect to 
independent variables, then perform the partial differentiations $\partial/\partial \tilde \psi$ and 
$\partial /\partial \tilde \phi$. The boundary conditions associated with the perturbed equations are
$\tilde \vvec(\pm 1,z,t)=0$ and $\tilde J_x(\pm 1,z,t)=0$. It is remarked that $\tilde J_z(\pm 1)$ can vary arbitrarily.

We consider $\tilde \psi(x,z,t)=\exp ({\rm i}kz-{\rm i}\omega t ) \Psi(x)$ and 
$\tilde \phi(x,z,t)=\exp({\rm i}kz-{\rm i}\omega t ) \Phi(x)$, where $\omega=\Omega +\zeta \, {\rm i}$ 
(${\rm i}=\sqrt{-1}$) and $\ell_z=2\pi/k$ is the wavelength of oscillations along the channel. 
$\Omega=2\pi/(\lambda t_B)=2\pi D/\lambda$ is the wave frequency and $\zeta$ is the growth/decay 
rate. Short-period transient oscillations with $\lambda \ll 1$ are dissolved by thermal fluctuations. 
Therefore, only long-period oscillations of $\lambda \gtrsim 1$ can exist. The linear solutions 
are decoupled in the $k$-space, and equations (\ref{eq:perturbed-psi-1}) and (\ref{eq:perturbed-psi-2}) 
remain invariant under the transformation $x\to -x$ if $\tilde \psi(x,z,t)=\mp \tilde \psi(-x,z,t)$ and 
$\tilde \phi(x,z,t)=\pm \tilde \phi(-x,z,t)$. This means that both symmetric and anti-symmetric 
modes are supported by the governing equations, and that degenerate pairs may exist in the 
eigenspectrum of $\omega$. Equation (\ref{eq:perturbed-psi-1}) reduces to Orr-Sommerfeld 
stability equation when $\phi=0$.

Most microchannels have typical widths of $2W\sim 10^{-4}$m. For $a\sim 10^{-6}$m, we will get 
$\epsilon \sim {\cal O}(10^{-5})$ and $0.01 \lesssim \Rey \lesssim 1$. On the other hand, the response 
time of $\tilde \phi$ to particle migrations is scaled by $\epsilon$, which is understood from 
equation (\ref{eq:perturbed-psi-2}). Therefore, the growth/decay rates of modes supported by 
diffusion, and not by the perturbations of streamlines, can be as small as $\zeta \sim {\cal O}(\epsilon) \sim 10^{-5}$, 
which requires a sophisticated numerical procedure to be resolved. 

We utilise the Chebyshev tau algorithm \citep{Orszag71,D96} to compute the wave functions $\Psi(x)$ 
and $\Phi(x)$, and assume $\Psi(x)=\sum p_n T_n(x)$ and $\Phi(x)=\sum q_n T_n(x)$ where $T_n(x)$ 
are Chebyshev polynomials defined over the region $-1 \le x \le +1$. There are six boundary conditions 
associated with $\tilde \psi$ and $\tilde \phi$. This suggests to assume six new unknowns, the so-called 
$\tau$ variables. Introducing a complex vector $\zvec$, which contains the variables $p_n$, $q_n$ 
($n=0,1,\cdots,N$) and $\tau_j$ ($j=1,2,\cdots,6$), and the Galerkin weighting of (\ref{eq:perturbed-psi-1}) 
and (\ref{eq:perturbed-psi-2}), leave us with the linear eigensystem $\Amat \cdot \zvec = \omega \Bmat \cdot \zvec$ 
where $\Amat$ and $\Bmat$ are complex matrices. In our implementation of the Chebyshev tau method, 
we use the formula $d^m T_n(x)/dx^m=2^{m-1}n (m-1)! \, C^{m}_{n-m}(x)$ \citep{GR07} where $C^{m}_{n-m}(x)$ 
are Gegenbauer polynomials. We find the generalized complex eigenvalues $\omega$ and their associated 
right eigenvectors using the routine zggev.f of LAPACK library. We begin our calculations with $N=20$ 
and increase $N$ until $\min [|\Omega|,|\zeta|]$ converges within $0.5\%$ for the mode with the smallest 
$|\zeta|$ in the spectrum. The major source of errors is the numerical evaluation of the inner products 
$\langle T_{n'},{\cal L}_{ij} T_n \rangle$, especially when the derivatives of $\phi_0(x)$ and $\Gamma_0(x)$ 
with jump discontinuities at $x=0$ appear in the integrand. We compute the inner products using a mid-point 
rule to avoid $x=0$, and use finer grids in the $x$-domain as $n$ or $n'$ increase. A uniform grid
helps us simultaneously resolve the strong near-wall features of certain modes and handle the central cusp. 
Reaching to $0.5\%$ error threshold occasionally needs $N\gtrsim 80$ to capture short wavelength modes 
when $\zeta \sim {\cal O}(\epsilon)$. To calibrate our code, we have solved the Orr-Sommerfeld stability equation 
and reproduced the results of \citet{Orszag71} up to the 8th decimal place. To assure that the physical 
eigenfrequencies are not sensitive to the choice of basis functions, we used Fourier series to reconstruct 
$\Phi(x)$ and $\Psi(x)$, and compared the resulting spectra with Chebyshev tau algorithm. The results of 
two methods match very well for modes that peak near the channel centreline. The Chebyshev tau method, 
however, gives more accurate results for modes that develop bumps near the walls. Our numerical 
calculations show that Fourier series increase the number of spurious modes.

\begin{table}
\begin{center}
\def~{\hphantom{0}}
  \begin{tabular}{lcccccccc}
   Mode                          &  $D$   &   ~$\Pen$  & $\Rey$ & $\phi_m \phi_{\rm bulk}$ &  $k$   & ~$\beta$ & $\omega^{\uparrow\downarrow}=\Omega+\zeta \, {\rm i}$ & $\omega^{\uparrow\uparrow}=\Omega+\zeta\, {\rm i}$ \\ \\
   ${\rm S}1,{\rm S}2$  & ~$0.084$      &   ~137 &  0.03 & 0.24  &  1.0    & ~5  & ~$0.017958-0.010418\,{\rm i}$  & ~$0.017886-0.010543\,{\rm i}$ \\  
   ${\rm S}3$                 & ~$0.084$      &    ~137  & 0.03 & 0.24 &  1.0    & ~5   & ~$0.482467-0.000397\,{\rm i}$  &                                                      \\ 
   ${\rm S}4$                 & ~$0.084$      &    ~137  & 0.03 & 0.24 &  1.0    & ~5  &                                                     & ~$0.471442-0.000425\,{\rm i}$  \\ \\
   ${\rm U}1,{\rm U}2$ & ~$0.030$  &   ~53.5   &  0.08 & 0.55  &  0.1   & ~2   & ~$0.000844+0.000282\,{\rm i}$  & ~$0.000851+0.000303\,{\rm i}$  \\  
   ${\rm U}3$                &  ~$0.030$  &   ~53.5   &  0.08 & 0.55 &  0.1   & ~2   &                                                     &  ~$0.006675+0.000053\,{\rm i}$ \\     
   \end{tabular}
  \caption{Eigenfrequencies of long-lived and unstable modes for a suspension flow with $K_c=0.03$ and $\epsilon=1.6 \times 10^{-5}$. 
  Degenerate pairs appear in a single row. The shape of each mode is identified through the symmetric ($^{\uparrow\uparrow}$) 
  or anti-symmetric ($^{\uparrow\downarrow}$) shape of $\Phi(x)$.}
  \label{table:eigvs}
  \end{center}
\end{table}

\begin{figure}
\centerline{\hbox{\includegraphics[width=0.45\textwidth]{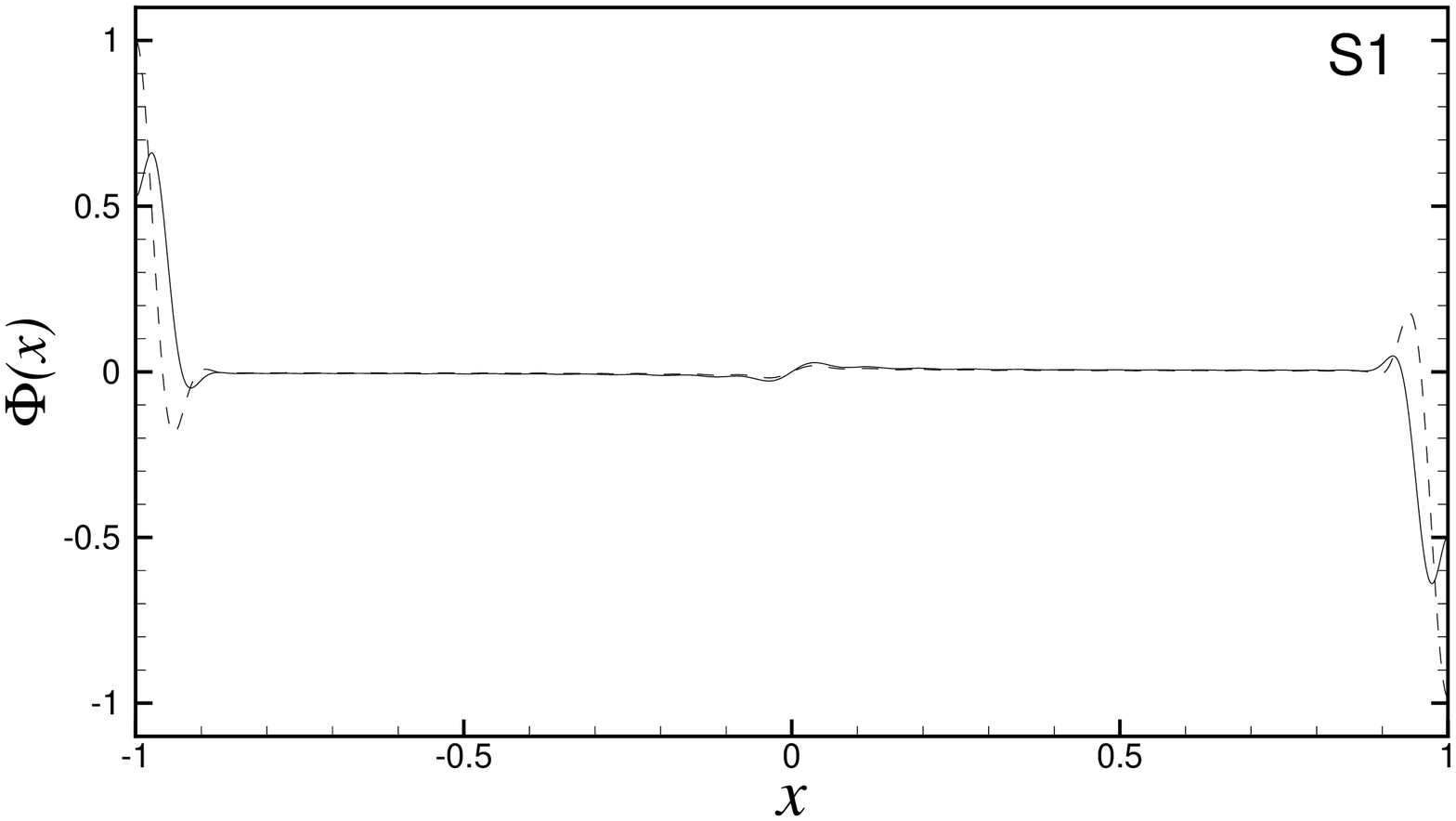}} 
             \hbox{\includegraphics[width=0.45\textwidth]{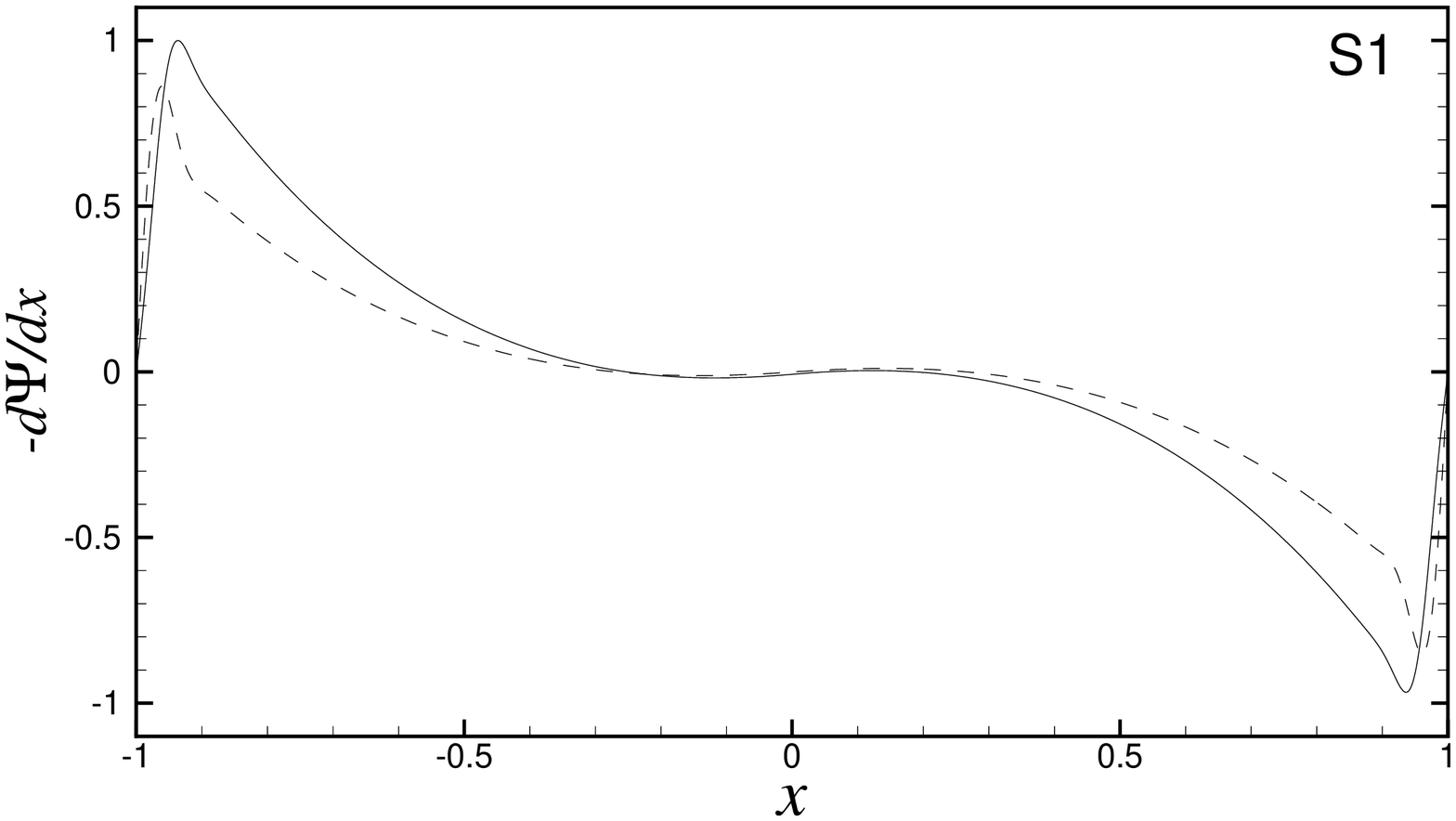} }	}
\centerline{\hbox{\includegraphics[width=0.45\textwidth]{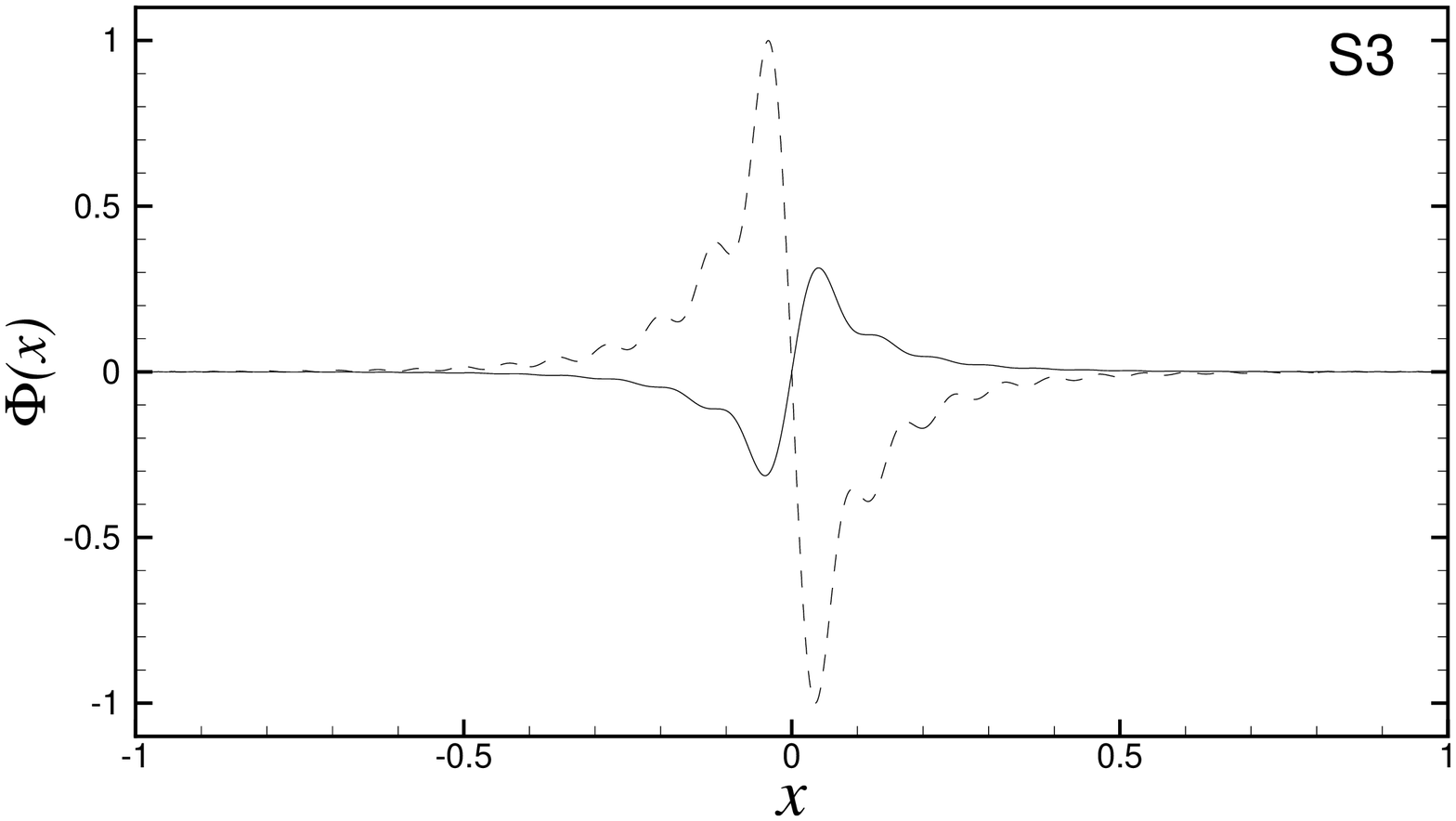}} 
             \hbox{\includegraphics[width=0.45\textwidth]{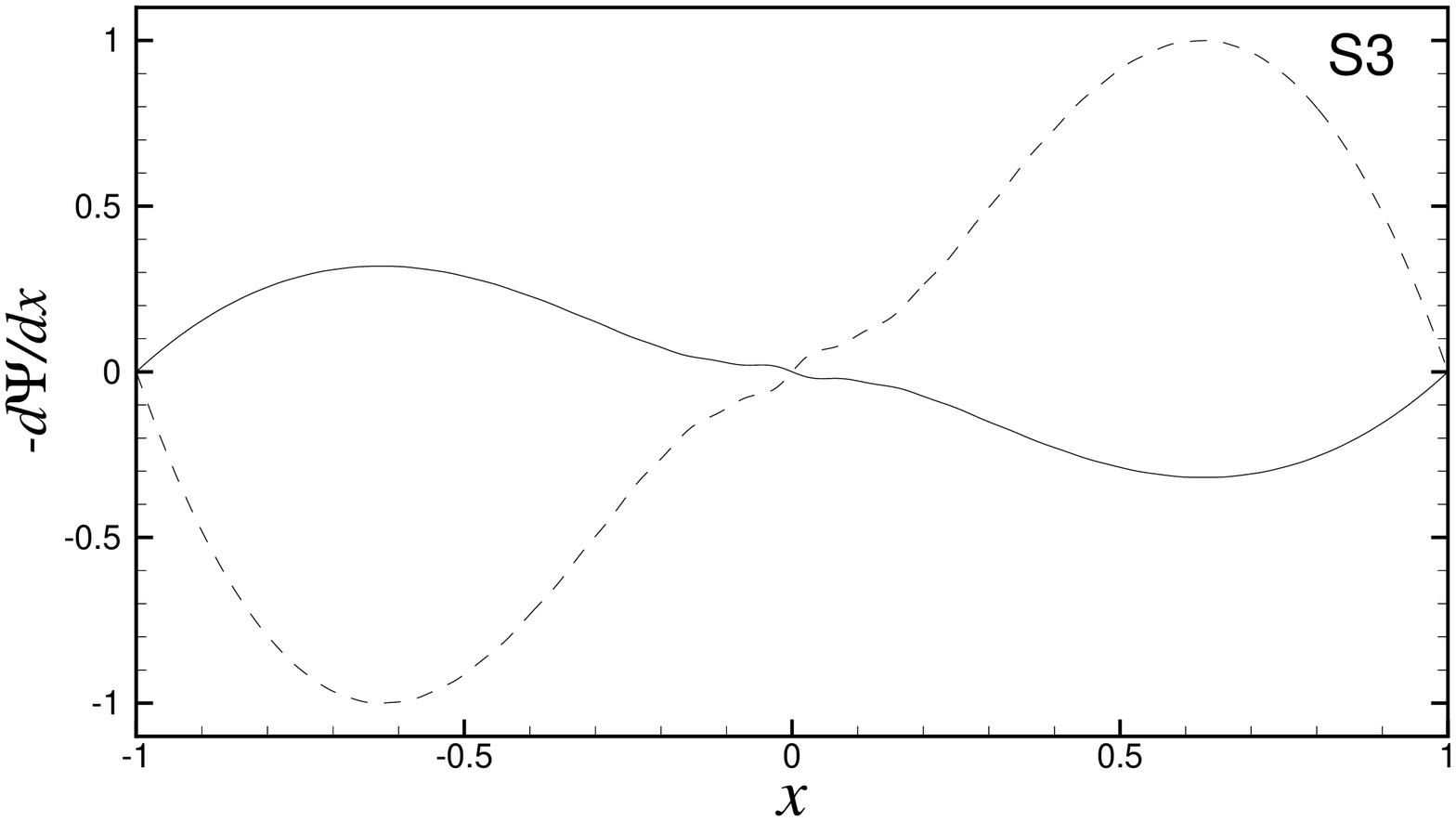} }	}      
\caption{The wave functions $\Phi(x)$ (left) and $-d\Psi(x)/dx$ (right) for the long-lived modes S1 
(top row) and S3 (bottom row). Note the strong near-wall features of mode S1. This property is
also shared by its symmetric partner S2. Solid and dashed lines correspond to the real and 
imaginary parts of the wave functions.}
\label{fig2}
\end{figure}

\section{Long-lived and unstable modes}
\label{sec:long-lived-and-unstable-modes}

We first carry out the stability analysis for the steady $\beta=5$ model of \S\ref{sec:steady-state}
(see figure \ref{fig1}). The main properties of this model, which we call model A,  are (i) relatively 
low $\phi_{\rm bulk}$; (ii) almost no flattening of the velocity profile near the channel centreline.
The matrices $\Amat$ and $\Bmat$ depend explicitly on $\epsilon$ and $\Rey$. In the limit $\epsilon = 0$,
the evolution of $\tilde \phi$ is only dictated by the deformations of streamlines, and we find only 
highly-damped, stable discrete modes ($\zeta \ll -1$), which are the characteristics of incompressible 
Newtonian flows at low Reynolds regimes. Turning on the effect of particle migration, $\epsilon \not =0$, 
gives birth to new long-lived modes with $-1 \ll \zeta < 0$. The eigenfrequencies of long-lived modes have 
been calculated for $k=1$ and given in Table \ref{table:eigvs}. These modes belong to two general 
families: degenerate and single modes. The oscillation periods and decay rates of modes S1 and S2 
are $\approx 26$ times larger than those of modes S3 and S4. We have plotted $\Phi(x)$ and the $x$-dependent 
part of $\tilde v_z(x,z,0) = -\exp({\rm i}kz) d\Psi(x)/dx$ in figure \ref{fig2} for modes S1 and S3. It is seen 
that the concentration profile of mode S1 has strong peaks near the walls while the dominant peaks 
of mode S3 have been generated near the channel centreline. Modes S1 and S2 have a better 
chance for being excited because particle-wall interactions can easily disturb the particle concentration 
and velocity field near the wall.  Modes S3 and S4 are most likely due to {\it collective} random motions 
because they have small periods of $\approx 1.1 \, t_B$ with $\Omega_{\rm S1}/\Omega_{\rm S3} \sim {\cal O}(D)$, 
decay slowly so that $\zeta_{{\rm S}_3}/\zeta_{{\rm S}_1} \sim {\cal O}(D)$, and have wide-spread patterns. 
S1 and S2 are therefore viscous modes supported by SID.

We now define the actual decay time of mode X as $t^{\rm X}_{d}=-(W/V_p) \ln(0.1)/ \zeta_{\rm X}$, which is 
the duration that the amplitude of $\tilde \phi$ decays to 10\% of its initial value. The actual oscillation period 
of mode X is given by $t^{\rm X}_{p}= (W/V_p) \lambda_{\rm X} t_B$. We find 
$(t^{\rm S1}_d,t^{\rm S1}_p)=(2.63\, {\rm s},4.16\, {\rm s})$. The characteristic times of mode S3 are quite 
surprising: $(t^{\rm S3}_d,t^{\rm S3}_p)=(69.05\, {\rm s},0.15\, {\rm s})$, which indicate an almost quasi-stationary 
oscillation in laboratory scales. All these modes will exhibit a high signal to noise ratio and can be measured 
by currently available high-speed imagers. The detection of S1 and S2 requires high spatial resolution, 
while S3 and S4 need high frame rates due to their small periods. The ripples in the concentration profiles, 
and the near-wall excess/deficit of particles observed in experiments \citep[see][]{Fra03,Sem07,Sem08,brown09} 
might be long-lived modes. Since $\zeta_{\rm S3}\approx \zeta_{\rm S4}$, we anticipate the coexistence 
of modes S3 and S4. The small difference between their frequencies can yield a quasi-periodic oscillation. 

\begin{figure}
\centerline{\hbox{\includegraphics[width=0.45\textwidth]{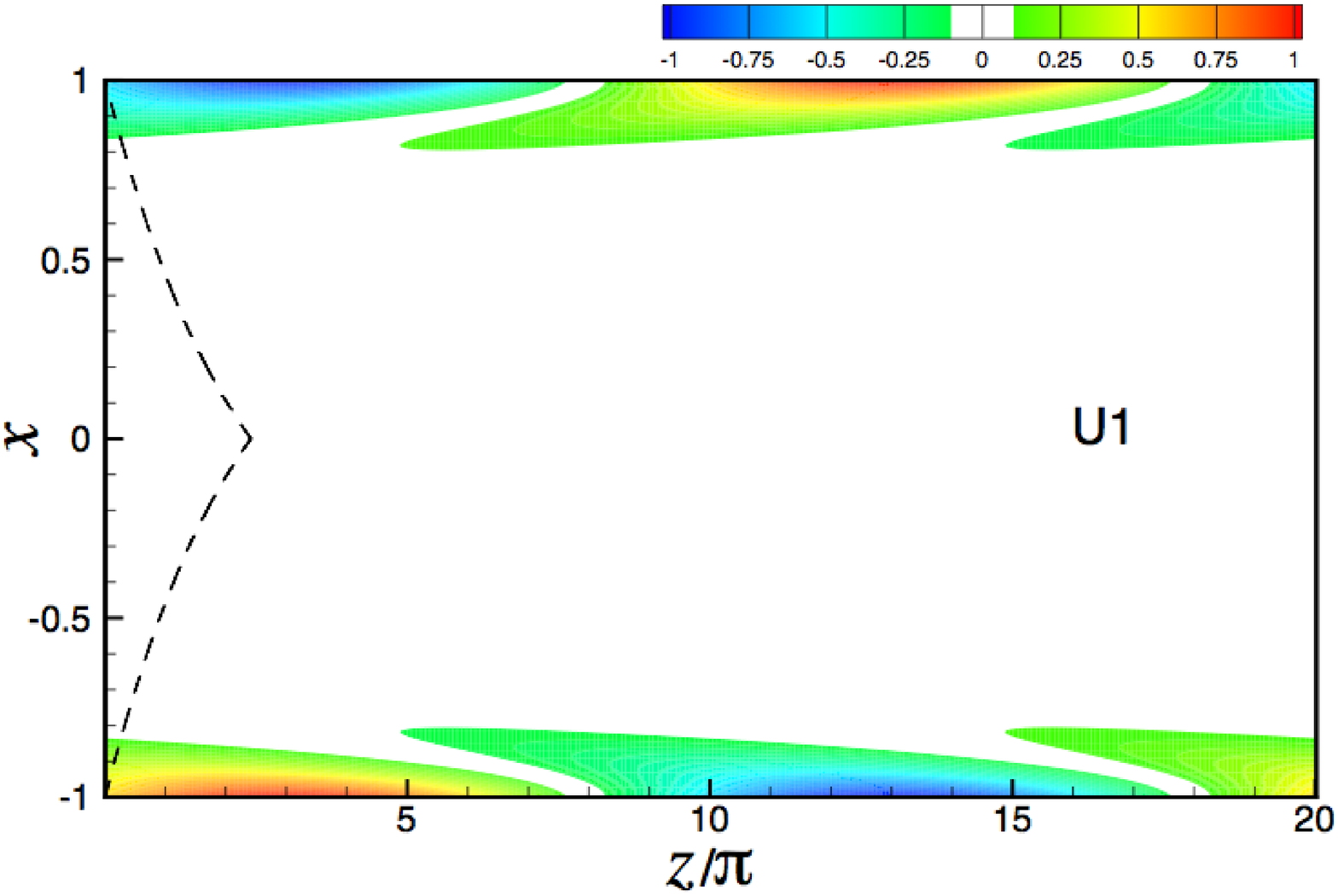}} 
             \hbox{\includegraphics[width=0.45\textwidth]{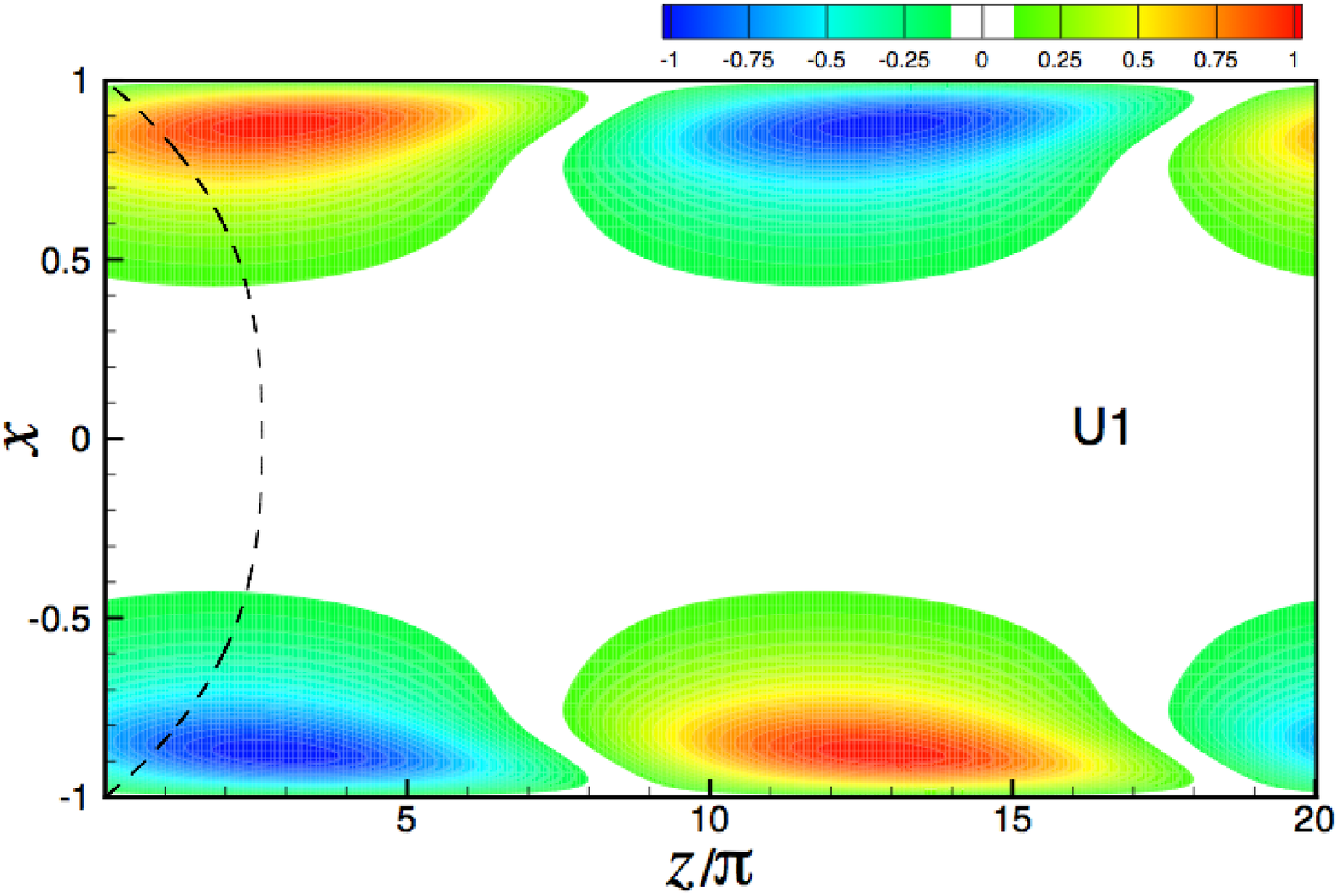} }	}
\centerline{\hbox{\includegraphics[width=0.45\textwidth]{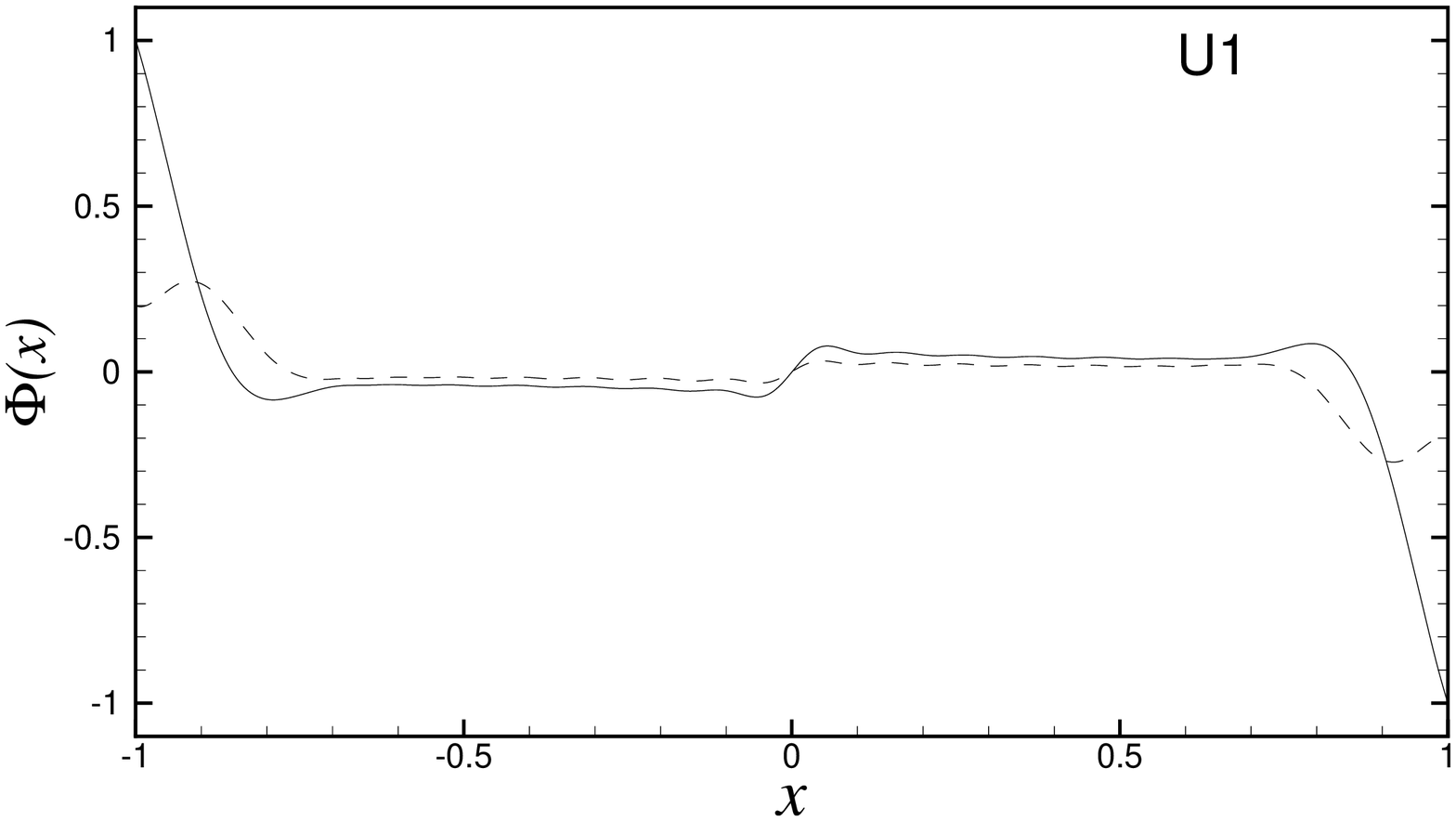}} 
             \hbox{\includegraphics[width=0.45\textwidth]{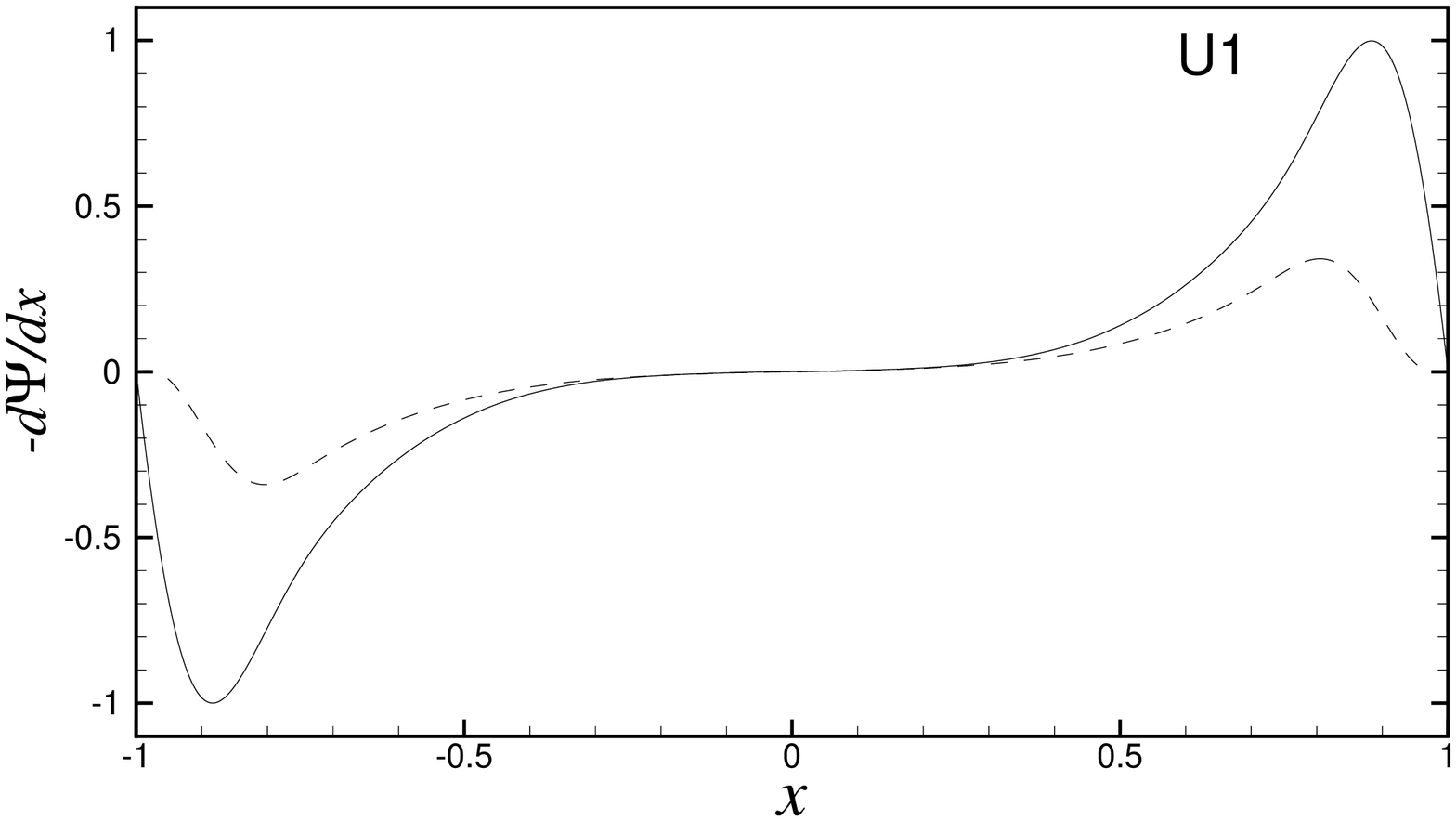} }	}
\centerline{\hbox{\includegraphics[width=0.45\textwidth]{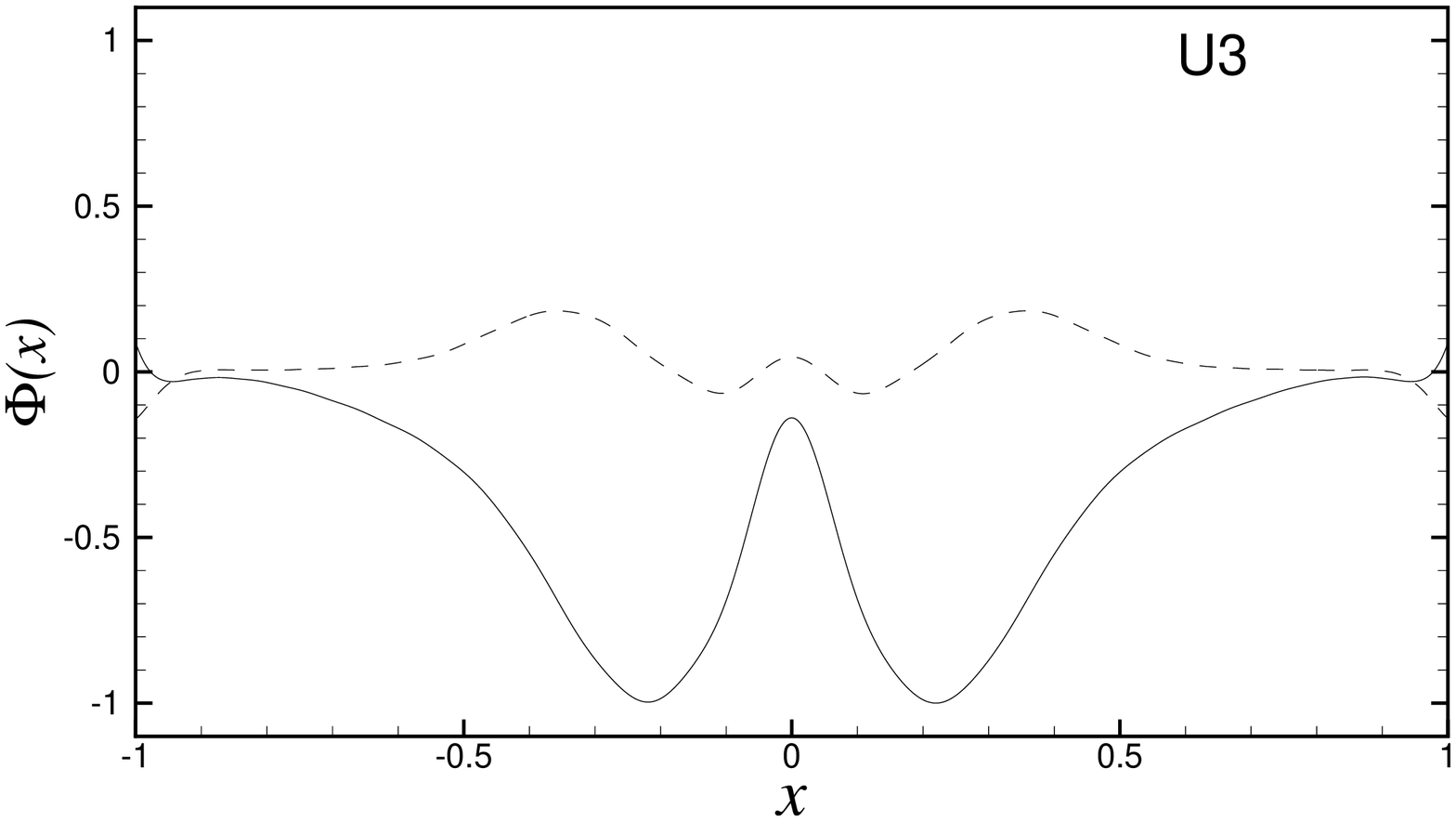}} 
             \hbox{\includegraphics[width=0.45\textwidth]{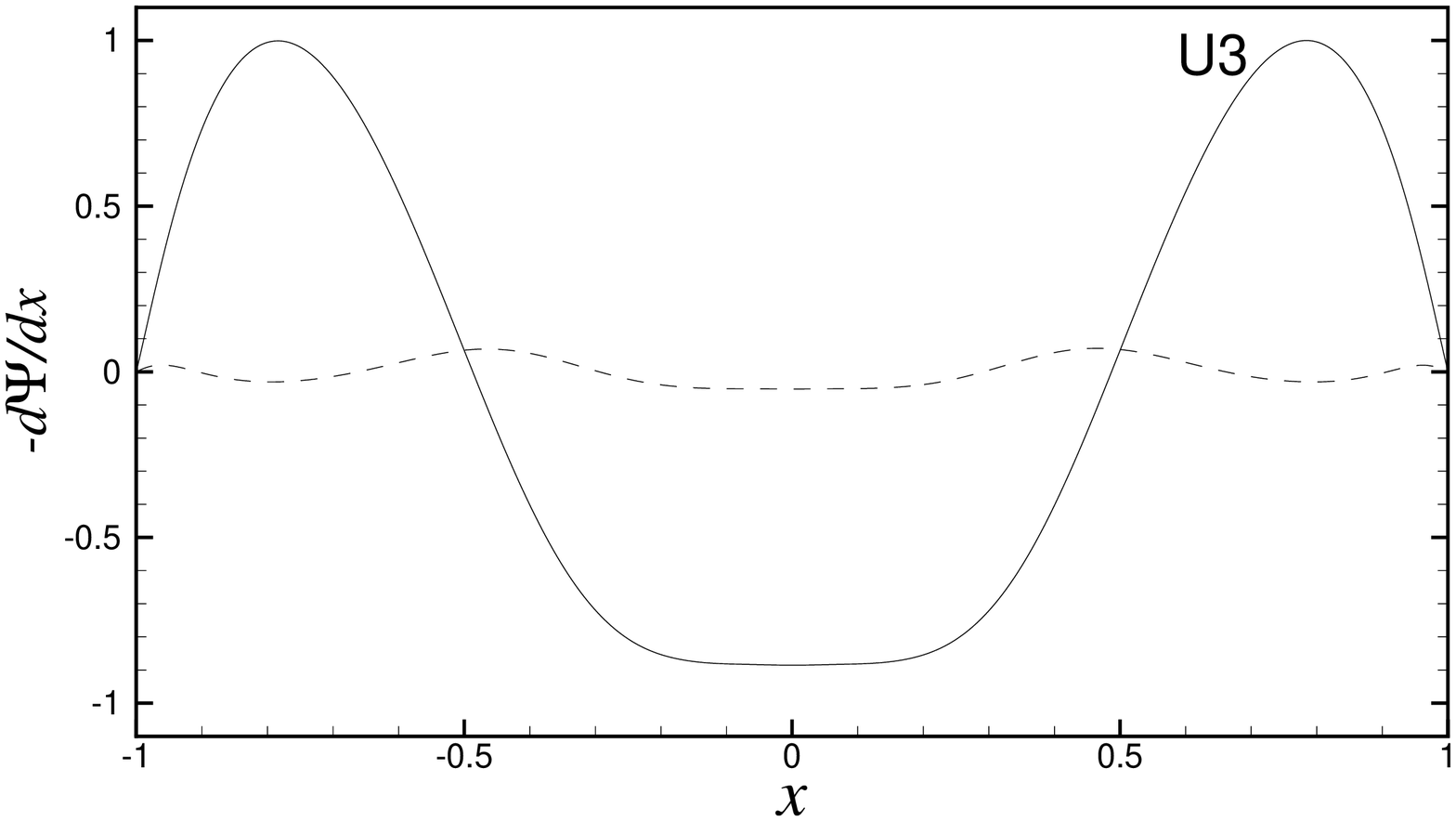} }	}             
\caption{The shapes of the unstable modes U1 and U3. Top panels demonstrate the 
contour plots of $\tilde \phi$ (left) and $\tilde v_z$ (right) at $t=0$ for mode U1.
Dashed lines in the panels of $\tilde \phi$ and $\tilde v_z$ show, respectively, the forms 
of the associated steady solutions $\phi_0(x)$ and $v_{z0}(x)$. The blank region corresponds 
to the lowest $10\%$ of the mode magnitude, which has not been shaded to highlight the major 
near-wall features. Middle and bottom panels show the wave functions $\Phi(x)$ (left) and 
$-d\Psi(x)/dx$ (right) for modes U1 and U3, respectively. Solid and dashed lines correspond 
to the real and imaginary parts of the wave functions.}
\label{fig3}
\end{figure}

Increasing $\phi_{\rm bulk}$ has a significant influence on the velocity profile and flattens it 
near the channel centreline \citep[\eg,][figure 4]{Sem07}. Our calculations show that in models 
with $\phi(0) < 1$ the flattening of the steady velocity curve is mainly controlled by the parameter 
$\beta$. The flattening of $v_{z0}(x)$ at $x=0$ is quantified by the curvature ${\cal C}=v_{z0,xx}(0)$,
which equals $-2$ for Poiseuille flow. To investigate the effect of ${\cal C}$ on the stability of 
suspension flows, we build another model B, and increase the average concentration to 
$\phi_{\rm bulk}=0.55/\phi_m$. We then set $V_p = 6\times 10^{-3}$m/s and $\beta=2$ to obtain a 
mass flow rate $Q=7\, {\rm nl/s}$. In this new flow regime, the curvature at $x=0$ becomes 
${\cal C}=-0.02235$ indicating a significant flattening. The perturbed equations now result in 
three unstable modes (U1, U2 and U3) with $\zeta > 0$. They have been reported in Table \ref{table:eigvs} 
for $k=0.1$ ($\ell_z=20\pi$). We have used this particular long wavelength because it expands 
the wavelengths of $\Phi(x)$ and $\Psi(x)$ in the $x$-direction, and leads to more visible patterns.

Unstable modes U1 and U2 are degenerate pairs, and U3 is a single symmetric mode. 
Modes U1 and U2 are the counterparts of S1 and S3. The oscillation periods of long 
wavelength unstable modes are: $t^{\rm U1}_p=31$, $t^{\rm U2}_p=30.76$ and $t^{\rm U3}_p=3.92$ 
seconds, and their amplitudes are magnified by a factor of 10 within $32$--$34$ seconds 
for the degenerate pair and $181.7$ seconds for mode U3. The dominant concentration 
and velocity peaks of modes U1 and U2 are developed near the channel wall (figure \ref{fig3}). 
The likelihood of exciting these modes is thus very high because of sedimentation or 
particle-wall interactions. In figure \ref{fig3}, we have also demonstrated the contours of 
$\Real[\tilde \phi]=\Real[\exp({\rm i}kz-{\rm i}\omega t)\, \Phi(x)]$ and 
$\Real[\tilde v_z]=\Real[-\exp({\rm i}kz-{\rm i}\omega t)\, d\Psi(x)/dx]$ for mode U1 at $t=0$.
Prominent curved tails have been developed in the wave packets of both $\tilde \phi$ and 
$\tilde v_z$. These features are shared by Kelvin-Helmholtz type instabilities that are usually 
triggered at interfaces, but our modes have emerged between the central region (where 
particles move with an almost constant velocity) and highly-sheared zone near the walls. 

The channel flow of suspensions with spherical particles involves five parameters ($\Rey$, 
$\Pen$, $\phi_0(0)$, $\beta$ and $D$) and it is impractical to survey the entire parameter space 
and identify its unstable zone. Here we vary only $\beta$, which also controls $\Pen$ and the 
normalized average concentration $\phi_{\rm bulk}$, and attempt to understand how the gradients 
of $\phi_0(x)$ and $v_{z0}(x)$ near the channel centreline (see \S\ref{sec:steady-state}) correlate 
with the instability. We find that by decreasing $\beta$ in model A, the magnitude of $\zeta$ drops for 
all modes until mode S3 becomes unstable for $\beta \lesssim 1.8$ ($\Pen \approx 124.3$, 
$\phi_{\rm bulk} \approx 0.29/\phi_m$). Other modes (S1, S2, and S4) are destabilized by further decreasing 
$\beta$. In another experiment, we gradually increased $\beta$ in the initially unstable model B to 
obtain more cuspy concentrations and less flattened velocity profiles as $|x|\to 0$. We find that mode 
U3 is stabilized for $\beta \gtrsim 4.5$ ($\Pen\approx 65.4$, $\phi_{\rm bulk} \approx 0.52/\phi_m$). 
Modes U1 and U2 resist stabilization until $\beta \approx 4.8$. It is seen that $\Pen$ should decrease
though mildly and $\phi_{\rm bulk}$ should increase to get instability. Nonetheless, both stable and 
unstable modes exist for $53 \lesssim \Pen \lesssim 124$, $0.42 \lesssim \phi_{\rm bulk} \lesssim 0.8$ and
$1.8 \lesssim \beta \lesssim 4.8$. The only property shared by all unstable systems is the flattened velocity 
profile, which constitutes a `necessary' condition for instability.  

Varying the wavelength $\ell_z=2\pi/k$ has also a notable impact on eigenfrequencies. 
Our calculations show that $\Omega$ and $|\zeta|$ increase proportional to $k$ in both the stable and 
unstable models. i.e., unstable modes grow faster for shorter wavelengths. Moreover, we find that the 
wavelengths of $\Phi(x)$ and $-d\Psi(x)/dx$ are approximately proportional to $\ell_z$, and consequently, 
the wave packets of degenerate modes become more compact near the walls for larger values of $k$ 
(compare the graphs of $-d\Psi(x)/dx$ in figures \ref{fig2} and \ref{fig3}). All these suggest that short 
wavelength instabilities can rapidly ruin the flow structure in the vicinity of the walls. Modes with longer 
wavelengths and lower oscillation frequencies can survive far from the walls and be observed experimentally. 
In a full three dimensional excitation with $\ell_y < \infty$, we anticipate faster growth and oscillation 
of unstable modes because clumps in the $y$-direction can probably enhance the migration 
of particles.

To this end, we argue that unstable modes U1 and U2 are amplified by the Brownian motion 
of particles. For a relatively low $\Pen$ and due to Brownian diffusion, some particles can `leak' 
from the region with flat velocity curve and move towards the channel walls. Such particles will 
locally increase the volume fraction and viscosity, while the suspension velocity drops. Note that 
the positive bumps of $\tilde \phi$ and $\tilde v_z$ are out of phase in the upper panels 
of figure \ref{fig3}. Particles initially moving close to the channel walls can also penetrate to inner 
regions through Brownian diffusion. When the overall velocity profile, $v_{z0}(x)+\tilde v_z(x,z,t)$, 
is flat near the channel centreline, SID cannot efficiently disperse particles trapped in the cavities 
of $(\tilde \phi,\tilde \vvec)$. Sustained Brownian migration, back and forth between highly-sheared 
and central regions, thus amplifies the concentration and velocity anomalies and triggers a mode. 
This instability mechanism works for mode U3 as well because the shape of $\tilde v_{z}$ is 
sufficiently flat in central regions (bottom panels in figure \ref{fig3}). Modes will be damped when 
SID becomes dominant due to the steady form of $\Gamma_0(x)$. In this condition SID will disperse 
particles participating in perturbations, and facilitate their migration towards the channel centre. 
This is why the transition from unstable to stable modes strongly correlates with the value of 
$\beta$, which controls the flatness of velocity curve.

\section{Conclusions}

We showed that the steady-state solutions of the model of \citet{Phil92} can reproduce recent 
experimental results of Brownian suspensions with spherical particles. We calculated the 
eigenmodes of the corresponding perturbed equations, and found new families of long-lived
and unstable modes. Unstable modes that we find occur when the fully-developed velocity 
profile is sufficiently flattened near the channel centreline. The fastest growing modes appear
as degenerate pairs, and they live inside the highly-sheared region near the walls. Since our
model has not taken particle-wall interactions into account, unstable modes seem to be triggered
by the transverse Brownian migrations across the streamlines of the fully developed flow. 
Mode amplification in models with flattened velocity profiles is mainly due to inefficient shear-induced
diffusion that can, in principle, disperse particles and force them to move towards the channel  
centreline. Rapidly growing unstable modes destruct the flow structure near the walls, and they 
can explain the experimentally observed excess and/or deficit of particles near the channel
walls. The instability mechanism that we suggested operates along the shortest direction  
of the channel. Therefore, dynamics along the neglected $y$-direction may only affect the 
wavelength and period of developing patterns, and not their general shapes. The three 
dimensional response of Brownian suspensions, especially in channels with $2W \sim H$, 
requires further exploration. \\

We thank Howard Stone for his insightful comments. A.K. thanks the department of Mechanical 
and Aerospace Engineering at Princeton University for their hospitality and generous support. 
We are indebted to anonymous referees whose criticisms helped us to substantially improve the 
presentation of the paper.


\begin{thebibliography}{99}

\bibitem[\protect\citeauthoryear{Brown et al.}{2009}]{brown09}
Brown, J. R., Fridjonsson, E. O., Seymour, J. D. \& Codd, S. L. 2009 
Nuclear magnetic resonance measurement of shear-induced particle migration in Brownian suspensions. 
\textit{Phys. of Fluids} \textbf{21}, 093301

\bibitem[\protect\citeauthoryear{Carpen \& Brady}{2002}]{Carpen02}
Carpen, I. C. \& Brady, J. F.  2002 Gravitational instability in suspension flow.
\textit{J. Fluid Mech.} \textbf{472}, 201-210

\bibitem[\protect\citeauthoryear{Dongarra et al.}{1996}]{D96}
Dongarra, J. J., Straughan, B. \& Walker, D. W.  1996 Chebyshev tau/QZ algorithm methods 
for calculating spectra of hydrodynamic stability problems. 
\textit{Journal of Applied Numerical Mathematics} {\bf 22}, 399-435

\bibitem[\protect\citeauthoryear{Frank et al.}{2003}]{Fra03}
Frank, M., Anderson, D., Weeks, E. R. \& Morris, J. F. 2003 Particle migration in pressure-driven
flow of a Brownian suspension. \textit{J. Fluid Mech.} \textbf{493}, 363-378

\bibitem[\protect\citeauthoryear{Govindarajan et al.}{2001}]{GNR01}
Govindarajan, R., Nott, P. R. \& Ramaswamy, S. 2001 Theory of suspension segregation in 
partially filled horizontal rotating cylinders. \textit{Phys. Fluids} \textbf{13}, 3517-3520

\bibitem[\protect\citeauthoryear{Gradshteyn \& Ryzhik}{2007}]{GR07}
Gradshteyn, I.S. \& Ryzhik, I.M. 2007 \textit{Table of Integrals, Series, and Products}.
Seventh Edition, Academic Press, Amsterdam

\bibitem[\protect\citeauthoryear{Helton \& Yager}{2007}]{Helton07}
Helton, K. L. \& Yager, P. 2007 Interfacial instabilities affect microfluidic extraction of 
small molecules from non-Newtonian fluids. \textit{Lab Chip} \textbf{7}, 1581-1588

\bibitem[\protect\citeauthoryear{Kauzlari\'c et al.}{2011}]{kau11}
Kauzlari\'c, D., Pastewka, L., Meyer, H., Heldele, R., Schulz, M., Weber, O., Piotter, V., 
Hausselt, J., Greiner, A. \&  Korvink, J. G. 2011 Smoothed particle hydrodynamics 
simulation of shear-induced powder migration in injection moulding.  
\textit{Phil. Trans. R. Soc. A} \textbf{369}, 2320-2328

\bibitem[\protect\citeauthoryear{Klinkenberg et al.}{2011}]{KLB11}
Klinkenberg, J., de Lange, H. C. \& Brandt, L. 2011 
Modal and non-modal stability of particle-laden channel flow. 
\textit{Phys. of Fluids} \textbf{23}, 064110

\bibitem[\protect\citeauthoryear{Kromkamp et al.}{2006}]{Kromkamp06}
Kromkamp, J., van der Padt, A., Schroen, C. G. P. H. \& Boom, R. M. 2006 Shear induced 
fractionation of particles. Patent EP1 673 957 A1, European Patent Office

\bibitem[\protect\citeauthoryear{Leighton \& Acrivos}{1987}]{Leighton87}
Leighton, D. \& Acrivos, A. 1987 The shear-induced migration of particles in concentrated suspensions. 
\textit{J. Fluid Mech.} \textbf{181}, 415-439

\bibitem[\protect\citeauthoryear{Merhi et al.}{2005}]{mer05}
Merhi, D., Lemaire, E., Bossis, G. \& Moukalled, F. 2005 Particle migration in a concentrated 
suspension flowing between rotating parallel plates: Investigation of diffusion flux coefficients. 
\textit{J. Rheol} \textbf{49}, 1429-1448

\bibitem[\protect\citeauthoryear{Miller \& Morris}{2006}]{MM06}
Miller, R. M. \& Morris J. F. 2006 Normal stress-driven migration and axial development 
in pressure-driven flow of concentrated suspensions. \textit{J. Non-Newtonian Fluid Mech.}
\textbf{135}, 149Ð165

\bibitem[\protect\citeauthoryear{Morris \& Boulay}{1999}]{MB99}
Morris, J. F. \& Boulay F. 1999 Curvilinear flows of noncolloidal suspensions: The role 
of normal stresses. \textit{J. Rheol.} \textbf{43}, 1213-1237

\bibitem[\protect\citeauthoryear{Nott \& Brady}{1994}]{not94}
Nott, P. R. \& Brady, J. F. 1994 Pressure-driven flow of suspensions: simulation and theory.
\textit{J. Fluid Mech.} \textbf{275}, 157-199

\bibitem[\protect\citeauthoryear{Orszag}{1971}]{Orszag71}
Orszag, S. A. 1971 Accurate solution of the Orr-Sommerfeld stability equation.
\textit{J . Fluid Mech.} \textbf{50}, 689-703

\bibitem[\protect\citeauthoryear{Phillips et al.}{1992}]{Phil92}
Phillips, R. J., Armstrong, R. C. \& Brown, R. A. 1992 A constitutive equation for concentrated 
suspensions that accounts for shear-induced particle migration. \textit{Phys. Fluids} \textbf{4}, 30-40

\bibitem[\protect\citeauthoryear{Rao et al.}{2007}]{Rao07}
Rao, R. R., Mondy, L. A. \& Altobelli, S. A. 2007 Instabilities during batch sedimentation in geometries 
containing obstacles: A numerical and experimental study. \textit{Int. J. Numer. Meth. Fluids} \textbf{55}, 
723-735

\bibitem[\protect\citeauthoryear{Rudyak et al.}{1997}]{R97}
Rudyak, V. Ya., Isakov, E. B. \& Bord, E. G. 1997 Hydrodynamic stability of the Poiseuille flow 
of dispersed fluid. \textit{J. Aerosol Sci.} \textbf{28}, 53-66

\bibitem[\protect\citeauthoryear{Rusconi \& Stone}{2008}]{RS08}
Rusconi, R. \& Stone, H. A. 2008 Shear-induced diffusion of platelike particles 
in microchannels. \textit{Phys. Rev. Lett.} {\bf 101}, 254502

\bibitem[\protect\citeauthoryear{Semwogerere et al.}{2007}]{Sem07}
Semwogerere, D., Morris, J. F. \& Weeks, E. R. 2007 Development of particle migration in pressure-driven 
flow of a Brownian suspension. \textit{J. Fluid Mech.} \textbf{581}, 437-451

\bibitem[\protect\citeauthoryear{Semwogerere \& Weeks}{2008}]{Sem08}
Semwogerere, D. \& Weeks, E. R. 2008 Shear-induced particle migration in binary colloidal suspensions.
\textit{Phys. Fluids} \textbf{20}, 043306

\bibitem[\protect\citeauthoryear{Stickel \& Powell}{2005}]{SP05}
Stickel, J. J. \& Powell, R. L. 2005 Fluid mechanics and rheology of dense suspensions.
\textit{Annual Review of Fluid Mechanics} \textbf{37}, 129-149

\bibitem[\protect\citeauthoryear{Vollebregt et al.}{2010}]{vol10}
Vollebregt, H. M., van der Sman, R. G. M. \& Boom, R. M. 2010 Suspension flow modelling in 
particle migration and microfiltration. \textit{Soft Matter} \textbf{6}, 6052-6064

\bibitem[\protect\citeauthoryear{Yurkovetsky \& Morris}{2008}]{YM08}
Yurkovetsky, Y. \& and Morris, J. F. 2008 Particle pressure in sheared 
Brownian suspensions. \textit{J. Rheol.} \textbf{52}, 141-164

\bibitem[\protect\citeauthoryear{Yiantsios}{2006}]{Yia06}
Yiantsios, S. G. 2006 Plane Poiseuille flow of a sedimenting suspension of Brownian 
hard-sphere particles: Hydrodynamic stability and direct numerical simulations. 
\textit{Phys. of Fluids} \textbf{18}, 054103


\end{thebibliography}
\end{document}